\documentclass[aps,prb,twocolumn,superscriptaddress,showpacs,showkeys]{revtex4-1}

\newcommand{\ped}[1]{\ensuremath{_{\rm #1}}}

\usepackage{graphicx}
\usepackage{amsmath}
\usepackage{xcolor}

\begin{document}

\title{Comprehensive Eliashberg analysis of microwave conductivity and penetration depth of K-, Co- and P-substituted BaFe$_2$As$_2$}

\author{D.~Torsello}
\affiliation{Politecnico di Torino, Department of Applied Science and Technology, Torino 10129, Italy}
\affiliation{Istituto Nazionale di Fisica Nucleare, Sezione di Torino, Torino 10125, Italy}
\author{G.~A.~Ummarino}
\affiliation{Politecnico di Torino, Department of Applied Science and Technology, Torino 10129, Italy}
\affiliation{National Research Nuclear University MEPhI (Moscow Engineering Physics Institute), Moskva 115409, Russia}
\author{L.~Gozzelino}
\affiliation{Politecnico di Torino, Department of Applied Science and Technology, Torino 10129, Italy}
\affiliation{Istituto Nazionale di Fisica Nucleare, Sezione di Torino, Torino 10125, Italy}
\author{T.~Tamegai}
\affiliation{The University of Tokyo, Department of Applied Physics, Hongo, Bunkyo-ku, Tokyo, 113-8656, Japan}
\author{G.~Ghigo}
\affiliation{Politecnico di Torino, Department of Applied Science and Technology, Torino 10129, Italy}
\affiliation{Istituto Nazionale di Fisica Nucleare, Sezione di Torino, Torino 10125, Italy}

\date{\today}

\begin{abstract}
We report on the combined experimental and theoretical analysis of microwave-frequency electromagnetic response of BaFe$_2$As$_2$ single crystals with different substitutions: K in the site of Ba (hole doping), Co in the Fe site (electron doping) and P in the As site (isovalent substitution). Measurements by a coplanar resonator technique lead to the experimental determination of the penetration depth and microwave conductivity as a function of temperature. The whole set of data is analyzed within a self-consistent three-band s$_\pm$-wave Eliashberg approach, able to account for all the main observed features in the different properties. Beside the validation of the model itself, the comparison between experiment and theory allows discussing the possible role of the Fe-As planes in defining the superconducting properties of these compounds, the relevance of coherence effects, and the presence of nodes in the superconducting order parameter.
\end{abstract}

\pacs{pacs}

\maketitle

\section{Introduction}
Iron-based superconductors (IBS) are among the most studied superconducting systems, due both to their potential for applications  in the high-field low-temperature regime \cite{Hosono2018,Pallecchi2015} and their interesting fundamental properties. Their discovery changed the idea that iron, with its large spin magnetic moment, had to be antagonistic against superconductivity and therefore should be avoided in the search for new superconducting materials and opened the way to the synthesis of a large variety of systems with different structure and parent materials \cite{Hosono2015}. Moreover, the discovery of antiferromagnetic order in the parent compounds of IBS \cite{Rotter2008}, as well as other more exotic effects such as the coexistence of ferromagnetism and superconductivity in Eu based systems \cite{Stolyarov2018}, generated interest in IBS as a playground to study the interplay between magnetism and superconductivity \cite{Dai2015,Stewart2011}. Among these materials, doped BaFe$_2$As$_2$ (Ba-122) compounds arose interest for the good quality of crystals available and also because superconductivity could be induced in many ways (by application of external pressure and by substitution of each atomic species) and resulting in phase diagrams that are very similar \cite{Canfield2010}.

The Ba-122 systems share the same building blocks with other IBS, with FeAs planes where Fe and As are in tetrahedral coordination. There is evidence that the transport, magnetic, and superconducting properties of these systems are controlled by such planes \cite{MazinPhysC}. The importance of the structural parameters as the As-Fe-As angle was outlined \cite{Garbarino2011}, but the FeAs planes also appear to tolerate considerable disorder, in contrast with the case of cuprates \cite{Sefat2008}.
Although these systems have been widely investigated in the last decade, theoretical approaches capable of capturing experimental data of different properties from different compounds in a self-consistent manner are still rare.

In this work, we show that the critical temperature, the quasiparticle conductivity at microwave frequencies, and the penetration depth of isovalent, hole- and electron-doped BaFe$_2$As$_2$ can be all understood in the framework of the same three-band, s$_\pm$-wave Eliashberg approach. With this aim, we analyze experimentally and with the same theoretical model, substitutions of all the species in the BaFe$_2$As$_2$ system, namely K in the site of Ba, Co in the Fe site, and P in the As site. The former is out of the FeAs planes, the latter two are in the FeAs planes. K substitution leads to a hole-doped superconductor, Co to an electron-doped one, while P is isovalent to As and exerts chemical pressure. As for the electronic structure, such Ba-122 compounds can be approximately described by three bands: two hole bands and one electron band for hole-doped and isovalent substitution (K and P), and two electron bands and one hole band for the electron-doped ones (Co)\cite{Ding2008,Terashima2009,Inosov}. Within the s$_\pm$-wave model,  coupling between the electron and the hole bands is due to the antiferromagnetic spin fluctuations, and the gaps of the electron bands have opposite sign with respect to the gaps of the hole bands.\cite{Mazin2008} The phononic contribution to the coupling is usually disregarded for these systems, but it was recently pointed out that it could play a non-negligible role \cite{wong2018}.

In the end, this combination of experimental data and theoretical insight enables us to discuss the possible origin of some observed features, such as the existence and origin of a peak in the quasiparticle conductivity, the difference of chemical substitution in or out of the FeAs planes, and the presence or absence of nodes in the superconducting gaps.

The paper is organized as follows. Details of the experimental techniques and the theoretical approach are given in Sect.II. They are the base for the comparison between measurements and calculations shown in Sect.III, in terms of penetration depth and microwave conductivity. In Sect.IV the outputs of the analysis are discussed and finally conclusions are drawn.

\section{Experimental techniques and theoretical methods}
\subsection{Preparation of the crystals}
\noindent Optimally doped single crystals of Ba$_{1-x}$K$_{x}$Fe$_{2}$As$_{2}$, Ba(Fe$_{1-x}$Co$_x$)$_{2}$As$_{2}$, and BaFe$_{2}$(As$_{1-x}$P$_x$)$_{2}$, with an analyzed doping level of x=0.42, 0.075, and 0.33, respectively, were grown by the FeAs self-flux method and some of their properties have been reported in Ref. \cite{Taen2013prb,Nakajima2010,Nakajima2012,Park2018}. All the investigated crystals were cleaved and reduced to the form of thin plates with thickness of about 10 $\mu$m, in the direction of the $c$-axis of the crystals, more than 10 times smaller than width and length.

\subsection{Microwave measurements}

\begin{figure}[t]
\begin{center}
\includegraphics[keepaspectratio, width=\columnwidth]{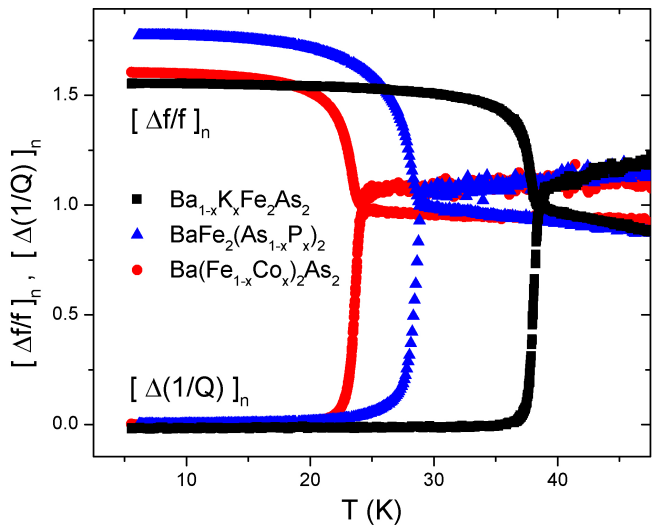}
\caption{Resonant frequency fractional shift and shift of the inverse of the quality factor are reported for the three investigated compounds. Data are normalized to the values at the critical temperature.}\label{fig_dati}
\end{center}
\end{figure}

\noindent Microwave measurements were carried out by means of a coplanar resonator technique, which has already been applied to characterize IBS thin crystals \cite{Ghigo2017prb,Ghigo2017scirep,Ghigo2018,Ghigo2018PRL}. In brief, the crystal under study is coupled to a coplanar-waveguide resonator, obtained by patterning an YBa$_2$Cu$_3$O$_{7-x}$ film on an MgO substrate \cite{Ghigo2012}. The crystal is placed far from edges, at the center of the stripline where the rf fields are uniform. Measurements of the resonance curves are repeated in the same conditions, with and without the crystal, by means of a vector network analyzer. A suitable input power is used, to ensure measurements are performed well below the non-linearity threshold for the resonator \cite{Ghigo2016}. Data are analyzed within a perturbative approach: shifts of the resonant frequency and changes of the unloaded quality factor relative to no sample coupled to the resonator (shown in Fig.\ref{fig_dati}) are related to the complex propagation constant, that in turn is related to the London penetration depth $\lambda_L$ and the quasiparticle conductivity $\sigma_1$ \cite{Ghigo2017prb}. The analysis procedure involves a self-consistent calibration, also accounting for the finite dimensions of the crystal, i.e. the penetration of the rf field also from the lateral sides of the sample.\cite{Ghigo2017prb}\\ Once $\lambda_L$ and $\sigma_1$ are obtained, they can be used to calculate the real and imaginary part of the surface impedance, $Z_s$, through
\begin{equation}
Z_{s}=R_{s}+iX_{s}=\frac{i\omega\mu_{0}\lambda_{L}}{\sqrt{1+i\omega\mu_{0}\sigma_{1}\lambda_{L}^2}}.
\label{eq:Zs}
\end{equation}

\begin{figure}[t]
\begin{center}
\includegraphics[keepaspectratio, width=\columnwidth]{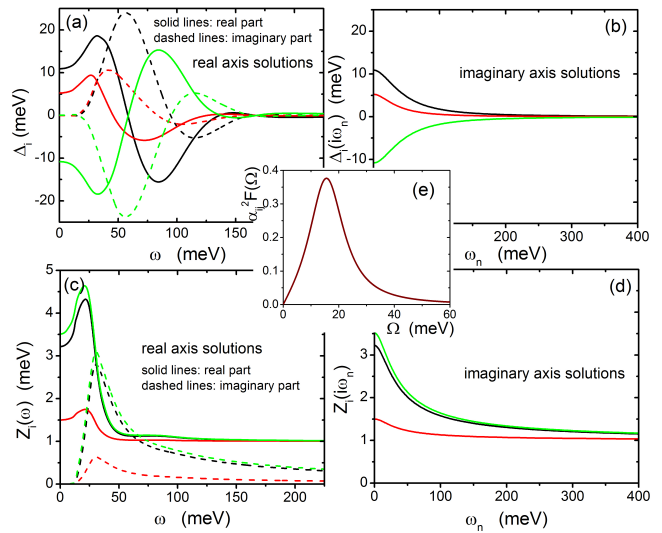}
\caption{Frequency dependence of the gaps obtained on the real axis (a) and imaginary axis (b) and of the renormalization frunctions  obtained on the real axis (c)  and imaginary axis(d). Black, red and green colors refer to bands 1 to 3. (e) Spectral function $\alpha_{ij}^2F$ for spin fluctuations in the superconducting state. All plots are given as an example for the K-doped system at T = 2K.}\label{Fig_theory}
\end{center}
\end{figure}

\begin{figure}[t]
\begin{center}
\includegraphics[keepaspectratio, width=\columnwidth]{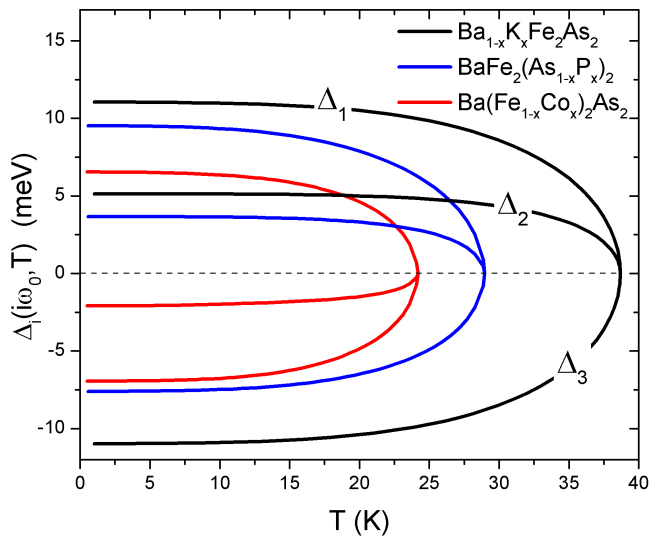}
\caption{Temperature dependence of the first value of the energy gaps for the three investigated compounds obtained by the solution of the imaginary-axis Eliashberg equations.}\label{Fig_Delta}
\end{center}
\end{figure}
\subsection{The Eliashberg approach}
\noindent In this work we aim at analyzing a rather large set of data within a comprehensive and self consistent theoretical approach. We chose to work with the Eliashberg equations that allow us to calculate all the measured properties starting from a choice of electron boson interaction and a description of the Fermi surface of the system. In the following we discuss the applicability of the model and the motivation for its use, as well as the approximations and assumptions made and why they apply to IBSs.\\
First of all,  it could be argued that a simpler BCS theory would also be a valid approach, but it is important to notice that BCS theory is not the Eliashberg weak coupling limit in a multiband system where intraband interaction is negligible (as it happens for IBSs): the two theories give quantitatively and qualitatively different results \cite{Dolgov2009PRB}. In Ref.\cite{Dolgov2009PRB} the authors also show that multi-band superconductivity is incorrectly described by the BCS formalism even for the weak-coupling limit unless a renormalization procedure is performed. A second observation can be made after performing Eliashberg calculations: the obtained renormalization functions largely deviate from the BCS value of 1 at low energy (see Fig.\ref{Fig_theory}), supporting the fact that BCS theory would be a too crude approximation for these materials.\\ 
Concerning approximations, our model does not include vertex corrections and makes use of the factorization of the momentum and energy dependence of the interaction propagator to obtain eq. \ref{eq:EE1} and \ref{eq:EE2}:  these are delicate points when the energy scale of the interaction is larger than the Fermi energy $E_F$.
In order to set the energy scales in the specific case of IBS, one should consider that the spin fluctuations, that provide the electron-boson coupling, in the superconducting state can be represented by a Lorentzian function \cite{Inosov} (see Fig.\ref{Fig_theory}(e)) peaked at $\Omega_0$. This energy scale follows the phenomenological law $\Omega_0($meV$)=2T_c($K$)/5$ \cite{Paglione2010} and is smaller than 15 meV for our systems, while the Fermi level is of the order of at least 100-200 meV \cite{Ortenzi2009} in IBSs. Therefore, although the two energy scales are not separated by severeal orders of magnitude as it happens in simpler systems, we are considering a regime in which the approximations described above  are valid. In particular, Migdal theorem (that ensures that one can neglect vertex corrections) is valid when $\lambda\Omega_0/E_F\ll1$, where $\lambda$ is the electron-boson coupling that is close to 1. The combination of these values ensures that the Migdal theorem holds and therefore that the Eliashberg theory is applicable to the present case.\\
However, to further support this choice other arguments can be brought based on the good agreement between measurements and calculations and on the equivalence of the obtainable results with more complex models.
Vertex corrections would in principle modify the expressions for the self energy and therefore change the Eliashberg equations (making them harder to treat numerically), yielding different gaps and renormalization functions from the ones we calculated (and that are in good agreement with literature). It was shown that in the first approximation these modifications only produce changes that can be emulated in a vertex-correction-free model with effettive coupling constant larger than its true value \cite{Pietronero1995,Grimaldi1995}. For this reason, together with the good overall agreement between calculations and experiments we prefer to neglect vertex corrections and work with a manageable theory.\\
Regarding how to treat the momentum and energy dependence of the interaction, an approach based on the anisotropic Eliashberg equation constructed avoiding the factorization of the momentum integration was proposed for MgB2 \cite{Choi2002} and gave good interpretation of experimental data. However, it was pointed out in a comment \cite{Mazin2004comment} that the same experimental data could be nicely reproduced also in a much simpler approach that makes use of momentum factorization. In IBSs, it has been found by ARPES measurements \cite{Terashima2009,Ding2008} that the gap amplitude on an individual Fermi surface sheet depends weakly on the direction. Although this is not a direct information about the momentum dependence of the interaction itself, this observation supports the choice of neglecting the momentum anisotropy of the interaction. In addition, the validity of this assumption is strongly supported by the very wide success obtained in fitting, interpreting and predicting experimental data with calculations based on these equations \cite{Parker2008,Popovich2010,Charnukha2011, Efremov2011,Tortello2010,Daghero2011}.

\subsection{Solving the Eliashberg equations}
\noindent In order to reproduce in a self-consistent way the whole set of experimental data and to estimate additional properties, we used a three-band Eliashberg s$_\pm$-wave model. The two (three, in the case of Co-doping) free parameters contained in the model (after reasonable assumptions are made, as explained in the next sections) have been fixed by the constraint that both the experimental $T_c$ and the temperature behaviour of the penetration depth, $\Delta\lambda_L(T)$, are simultaneously reproduced.
To calculate the gaps and the critical temperature by the three-band Eliashberg equations \cite{Eliashberg,chubukov2008,bennemann2008superconductivity} we solve six coupled equations for the frequency dependent gaps $\Delta_{i}(i\omega_{n})$ and renormalization functions $Z_{i}(i\omega_{n})$, where $i$ is a band index ranging from $1$ to $3$ and $\omega_{n}$ are the Matsubara frequencies. The imaginary-axis equations \cite{Eliashberg,Korshunov,UmmarinoPRB2011} read: 
\begin{eqnarray}
&&\omega_{n}Z_{i}(i\omega_{n})=\omega_{n}+ \pi T\sum_{m,j}\Lambda^{Z}_{ij}(i\omega_{n},i\omega_{m})N^{Z}_{j}(i\omega_{m})\nonumber\\
&&+\sum_{j}[\Gamma^{N}\ped{ij}+\Gamma^{M}\ped{ij}]N^{Z}_{j}(i\omega_{n})
\label{eq:EE1}
\end{eqnarray}
\begin{eqnarray}
&&Z_{i}(i\omega_{n})\Delta_{i}(i\omega_{n})=\pi
T\sum_{m,j}\big[\Lambda^{\Delta}_{ij}(i\omega_{n},i\omega_{m})-\mu^{*}_{ij}(\omega_{c})\big]\nonumber\\
&&\times\Theta(\omega_{c}-|\omega_{m}|)N^{\Delta}_{j}(i\omega_{m})
+\sum_{j}[\Gamma^{N}\ped{ij}+\Gamma^{M}\ped{ij}]N^{\Delta}_{j}(i\omega_{n}),\phantom{aaa}
 \label{eq:EE2}
\end{eqnarray}
where $\Gamma^{N}\ped{ij}$ and $\Gamma^{M}\ped{ij}$ are the scattering rates from non-magnetic and magnetic impurities, $\Lambda^{Z}_{ij}(i\omega_{n},i\omega_{m})=\Lambda^{ph}_{ij}(i\omega_{n},i\omega_{m})+\Lambda^{sf}_{ij}(i\omega_{n},i\omega_{m})$ and
$\Lambda^{\Delta}_{ij}(i\omega_{n},i\omega_{m})=\Lambda^{ph}_{ij}(i\omega_{n},i\omega_{m})-\Lambda^{sf}_{ij}(i\omega_{n},i\omega_{m})$, where
\begin{eqnarray}
&&\Lambda^{ph,sf}_{ij}(i\omega_{n},i\omega_{m})=\notag\\
&&=2\int_{0}^{+\infty}d\Omega \Omega\alpha^{2}_{ij}F^{ph,sf}(\Omega)/[(\omega_{n}-\omega_{m})^{2}+\Omega^{2}].
\end{eqnarray}
\noindent $\Theta$ is the Heaviside function and $\omega_{c}$ is a cutoff
energy. The superscripts $ph$ and $sf$ indicate, respectively, the phonon and spin-fluctuations terms of the frequency dependent spectral functions $\alpha^{2}_{ij}F(\Omega)$, considered to have a Lorentzian shape (see Fig.\ref{Fig_theory}(e)):
\begin{equation}
\alpha_{ij}^2F^{sf}(\Omega)= C_{ij}\big\{L(\Omega+\Omega_{ij},Y_{ij})-
L(\Omega-\Omega_{ij},Y_{ij})\big\},\notag
\end{equation}
where
\[
L(\Omega\pm\Omega_{ij},Y_{ij})=\frac{1}{(\Omega \pm\Omega_{ij})^2+Y_{ij}^2}
\]
and $C_{ij}$ are normalization constants, necessary to obtain the proper values of $\lambda_{ij}$, while $\Omega_{ij}$ and $Y_{ij}$ are the peak energies and the half-widths of the Lorentzian functions, set to be $\Omega_{ij}=\Omega_{0}$ and  $Y_{ij}= \Omega_{0}/2$, based on the results of inelastic neutron scattering measurements \cite{Inosov}.
The quantities $\mu^{*}_{ij}(\omega\ped{c})$ are the elements of the $3\times 3$
Coulomb pseudopotential matrix. Moreover,
$N^{\Delta}_{j}(i\omega_{m})=\Delta_{j}(i\omega_{m})/
{\sqrt{\omega^{2}_{m}+\Delta^{2}_{j}(i\omega_{m})}}$ and
$N^{Z}_{j}(i\omega_{m})=\omega_{m}/{\sqrt{\omega^{2}_{m}+\Delta^{2}_{j}(i\omega_{m})}}$.
Finally, the electron-boson coupling constants are defined as
$\lambda^{ph,sf}_{ij}=2\int_{0}^{+\infty}d\Omega\frac{\alpha^{2}_{ij}F^{ph,sf}(\Omega)}{\Omega}$.

The gaps are assumed to be isotropic due to the low values of anisotropy typical of optimally doped Ba-122 compounds. Moreover, considering gap anisotropy would greatly complicate the equations and make the comparison with the experiment unpractical, without significantly changing the physics of these systems.

The solution of Eqs. \ref{eq:EE1} and \ref{eq:EE2} requires a large number of input parameters. In part they can be taken or deduced from the literature, in part they can be fixed by reasonable assumptions and approximations (such as setting to zero the impurity scattering rates and Coulomb pseudopotential matrix elements), see Ref.$\cite{Ghigo2017prb}$ for a detailed discussion. The remaining free parameters (basically the non-zero $\lambda_{ij}$ values reported in Table \ref{tab:parametri}) are then adjusted to reproduce the experimental data at best. Figure \ref{Fig_Delta} shows, for the three compounds, the temperature dependence of the first value of the energy gaps obtained by the solution of the imaginary-axis Eliashberg equations, that are the basis for the calculation of the penetration depth.

\subsection{Penetration depth calculation}
\noindent 
The penetration depth can be computed starting from the gaps $\Delta_{i}(i\omega_{n})$ and the renormalization functions $Z_{i}(i\omega_{n})$ by
\begin{eqnarray}
&&\lambda^{-2}(T)=(\frac{\omega_{p}}{c})^{2}
\sum_{i=1}^{3}w^\lambda_i\pi T\notag\\
&&\times\sum_{n=-\infty}^{+\infty}\frac{\Delta_{i}^{2}(\omega_{n})Z_{i}^{2}(\omega_{n})}{[\omega^{2}_{n}Z_{i}^{2}(\omega_{n})+\Delta_{i}^{2}(\omega_{n})Z_{i}^{2}(\omega_{n})]^{3/2}}\label{eq.lambda}
\end{eqnarray}
where $w^\lambda_i=\left(\omega_{p,i}/\omega_{p}\right)^{2}$ are the weights of the single bands, $\omega_{p,i}$ is the plasma frequency of the $i$-th band and $\omega_{p}$ is the total plasma frequency. Here, we can only act on the weights $w^\lambda_i$ in order to adapt the calculation to the experimental $\lambda_L(T)$. The multiplicative factor that involves the plasma frequencies derives from the fact that the low-temperature value of the penetration depth $\lambda_L(0)$ should, in principle, be related to the plasma frequency by $\omega_p=c/\lambda_L(0)$. This is strictly valid only for a clean uniform superconductor at $T=0$ if strong-coupling effects (or, more generally, Fermi-liquid effects) are negligible. \\
It should be noted that eq. \ref{eq.lambda}, does not include cross terms but only single-band ones, representing the contribution to the superfluid density of each specific band. Cross-terms would stem from Cooper pairs composed of electrons located on different bands. Since they would be characterized by very different momentum $k$ the probability of forming such a pair is vanishing small and therefore the cross terms can be neglected.

\subsection{Conductivity calculation}
The solutions of the Eliashberg equations can also be used to determine the microwave conductivity. In this case, it is more convenient to work with the real-axis formalism of the same model, which is equivalent, being frequency dependent and using exactly the same input parameters). The conductivity is then calculated from \cite{Dolgov2009}
\begin{eqnarray}
&&\sigma_1(\omega\rightarrow0)=\sum_{i}w^\sigma_{i}\sigma_{1,i}=\sum_{i}w^\sigma_{i}A_i\label{eq.cond}\\
&&\times\int_{0}^{+\infty}d\omega\left(-\frac{\partial f(\omega)}{d\omega}\right)\left[(\operatorname{Re}g_i^Z(\omega))^2+(\operatorname{Re}g_i^\Delta(\omega))^2\right]\notag
\end{eqnarray}
\noindent where $i$ is the band index, $A_i=N_i(0)v_F^2e^2\tau_i(T)$, $w^\sigma_{i}$ is the weight of the $i$-th band, and 
\begin{eqnarray}
&&g_i^Z(\omega)= Z_i(\omega)\omega/\sqrt{[Z_i(\omega)\omega]^2-[\Delta_i^2(\omega)Z_i^2(\omega)]}\notag\\
&&g_i^\Delta(\omega)= \Delta_i(\omega)Z_i(\omega)/\sqrt{[Z_i(\omega)\omega]^2-[\Delta_i^2(\omega)Z_i^2(\omega)]}\notag
\end{eqnarray}

The density of states, $N_i(0)$, and the scattering time, $\tau_i(T)$, for each band are unknown for these compounds. An estimation of an effective scattering time can be obtained by means of the phenomenological two-fluid model (see below). Then, assuming that $\tau_i$ has the same temperature dependence for all the bands, and all scale factors $A_i$ are equal ($=A$), the weights in Eq.\ref{eq.cond} can be tuned to reproduce the experimental data.

\subsection{Determination of an effective scattering time within a two-fluid model}
\begin{figure}[ht]
\begin{center}
\includegraphics[keepaspectratio, width=\columnwidth]{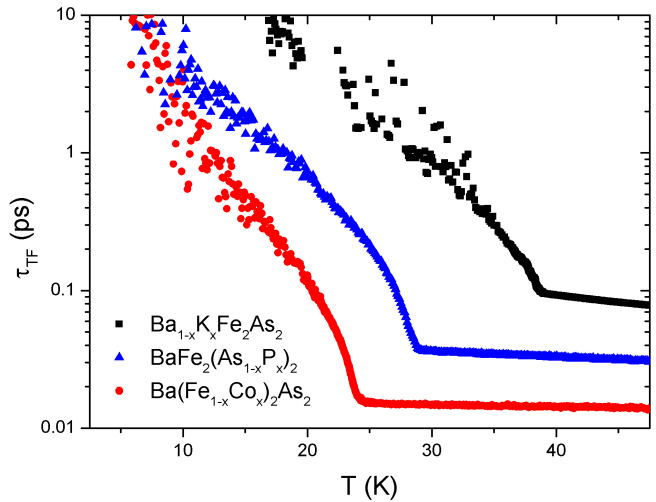}
\caption{Scattering time obtained by the surface impedance data through a two-fluid model.}\label{fig_tau}
\end{center}
\end{figure}

\noindent The two-fluid model provides a very useful basis for a first understanding of the role of inelastic scattering in these compounds \citep{Ghigo2018}. It gives the way of extracting a temperature-dependent scattering time $\tau_{TF}$($T$) from the experimental surface impedance. In the standard model, the surface impedance reads
\begin{equation}
Z_s=R_s+i X_s = \sqrt{i\mu_0 \omega/(\sigma_1-i\sigma_2)}\notag
\end{equation}
and in turn the conductivity can be expressed, in the limit $\omega\tau\ll1$, as
\begin{equation}
\sigma_1=2\omega\mu_0\frac{R_sX_s}{(R_s^2+X_s^2)^2}\label{eq_s1}
\end{equation}
\begin{equation}
\sigma_2=\omega\mu_0\frac{X_s^2-R_s^2}{(R_s^2+X_s^2)^2}\label{eq_s2}.
\end{equation}
Assuming that the conductivity of the normal fluid can be modeled by a Drude form, and $n_s(0)=n_s(T)+n_n(T)$, where $n_s$ and $n_n$ are the superfluid and quasiparticle densities, respectively, and considering the London relation $n_s(T)=m^*/(\mu_0e^2\lambda_L^2(T))$, the complex conductivity can also be written as
\begin{equation}
\sigma_1-i\sigma_2=\frac{n_ne^2}{m^*}\frac{\tau_{TF}}{1+i\omega\tau_{TF}}-\frac{i}{\mu_0\omega\lambda_L^2(T)}.\label{eq_Drude}
\end{equation}
Then, combining Eqs. \ref{eq_s1}$-$\ref{eq_Drude}, it turns out that
\begin{equation}
\tau_{TF}^{-1}=\frac{1}{\mu_0\lambda_L^2(0)\sigma_1}-\frac{\omega(X_s^2-R_s^2)}{2X_sR_s}.\label{eq_tau}
\end{equation}
Therefore $\tau_{TF}$ can be calculated from the microwave measurements experimental data. The values of $\tau_{TF}$ obtained by Eq.\ref{eq_tau} are affected by high uncertainty at temperatures below $T_c$/2. They are shown in Fig.\ref{fig_tau} and used as a base to define, by a proper smoothing procedure, the $\tau(T)$ in Eq.\ref{eq.cond}. The value of $\tau_{TF}(40 K)\approx 0.1$ ps for the Ba$_{1-x}$K$_{x}$Fe$_{2}$As$_{2}$ crystal is in good agreement with previous results \cite{Hashimoto2009PRL2}. At low temperatures, far below $T_c$, the quasiparticle scattering time reaches 10 ps and more, values that are more than two orders of magnitude larger than that in the normal state, as it happens also for other IBS systems \cite{Takahashi2011}.

\section{Comparison between experimental data and calculations}
\begin{figure}[ht]
\begin{center}
\includegraphics[keepaspectratio, width=\columnwidth]{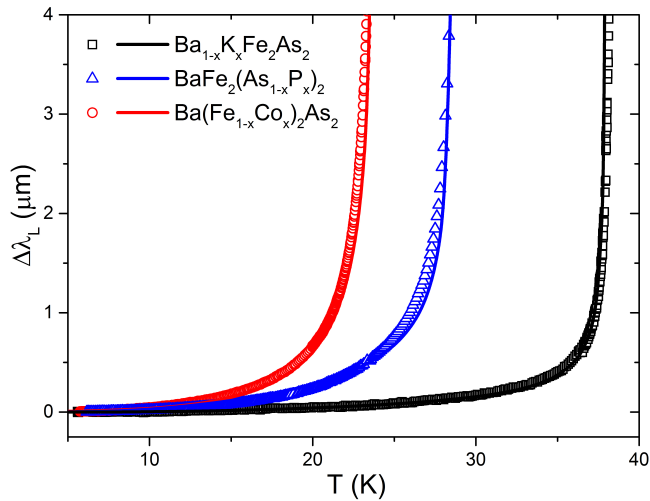}
\caption{Comparison between experimental (symbols) and calculated (lines) penetration depth increments, $\Delta\lambda_L=\lambda_L(T)-\lambda_L(0)$, for the three compounds.}\label{Fig_lambda}
\end{center}
\end{figure}

\textit{(a) Penetration depth} $-$
Figure \ref{Fig_lambda} shows the experimental penetration depth increments $\Delta\lambda_L=\lambda_L(T)-\lambda_L(0)$ for the three investigated compounds (symbols). Data are compared to calculations (lines) performed with the approach described in Sects. IIC and IID, with the parameters that are reported in Table \ref{tab:parametri} and that yield the gaps in Fig.\ref{Fig_Delta}. The parameters values were chosen to give the best agreement with the experimental curves, within the very limited range of variability allowed by all the constraints assumed in the model. Among them, we imposed that the low-temperature values of the gaps must be comparable with those reported in literature for ARPES measurements \cite{Ding2008,Terashima2009,Inosov2011,Mansart2012,Yoshida2014,Ye2012} (see real-axis values in Table \ref{tab:parametri}).

In all cases, the agreement between the experimental and the theoretical $\Delta\lambda_L(T)$ curves is excellent, but a distinction should be made between the K-doped case and the other ones. In the former, with a value of $\omega_{p}$=1 eV in good agreement with literature data, it has been possible to obtain by the model a $\lambda_L(0)$ value of  230 nm, in remarkable agreement with the experimental one, i.e. $\lambda_L(0)$=197 nm. 
On the contrary, for the Co- and P-substituted systems the comparison is more problematic. Still the temperature dependence of $\lambda_L(T)$ is quite well reproduced, but the theoretical $\lambda_L(0)$ values are much larger that the experimental one (even up to a factor of 10 for Co doping). This has already been reported for Co-doped samples \cite{Vorontsov2009} and will be discussed in the next section.
In addition, it should also be noted that, different from the case of K- and P- doping, in the Eliashberg calculations for Co-doping it was necessary to include a non-negligible intraband (phononic) coupling contribution\cite{wong2018} (coupling constants $\lambda_{ii}$=0.3).

\textit{(b) Microwave conductivity} $-$ The same input parameters optimized to reproduce $T_c$ and $\Delta\lambda_L(T)$ were then used to compute the quasiparticle conductivity, as described above. It is noteworthy that, at this point, the only free parameters left were the weights of the bands, $w^\sigma_i$ (reported in Table \ref{tab:parametri}), and the scale factor, $A$. Figure \ref{Fig_sigma} shows for each compound the comparison between the experimental and the calculated conductivity. It turns out that the temperature position of the broad peak below, but close to, $T_c/2$ is fairly well reproduced for all the compounds. The curves are in good agreement for P- and Co-dopings, less good but still reasonable for the K-substituted sample. All these experimental findings and calculations are comprehensively discussed in the next Section.

\section{Discussion and conclusions}
As already stated, this study represents an attempt to coherently interpret, within the same theoretical model, multiple properties measured experimentally on samples of the same system but with different types of doping. It was shown above that the measured critical temperature and the temperature dependence of the penetration depth and quasiparticle conductivity are, on the whole, very satisfactorily reproduced within a three band $s_{\pm}$ Eliashberg model, also yielding gap values in good agreement with ARPES measurements\cite{Ding2008,Inosov2011,Mansart2012,Yoshida2014,Ye2012}. Indeed, it has to be stressed that the constraints assumed for the calculations and the requirement to reproduce so different experimental outcomes reduced a lot the number of free parameters and their range of variability. Therefore, even if in some cases the quantitative matching is not perfect, the overall agreement with the experiment should be considered remarkable and fully successful. In this Section, the main results reported previously are discussed in the frame of the present literature, to point out similarities and discrepancies, and to draw reasonable conclusion.

\begin{figure}[ht]
\begin{center}
\includegraphics[keepaspectratio, width=\columnwidth]{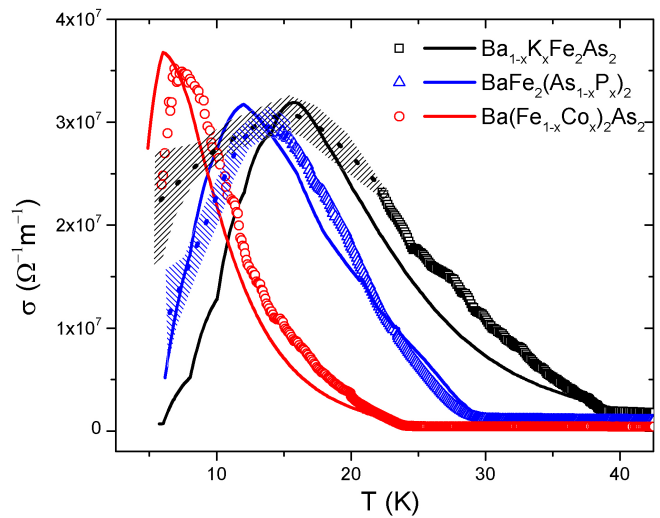}
\caption{Comparison between experimental (symbols) and calculated (lines) quasiparticle conductivity, for the three compounds. A smoothing of the low-temperature experimental points has been done in some cases (dotted lines) due to the high uncertainty of data (shaded areas correspond to uncertainty regions).}\label{Fig_sigma}
\end{center}
\end{figure}

\textit{(a) Critical temperature and energy gaps} $-$ The experimental critical temperature has been defined as the temperature at which $\lambda_L(T)$ diverges, and is listed in Table \ref{tab:parametri} for the three compounds under study. $T_c$ values are in accordance with literature \cite{Korshunov2017,Inosov2011}, for the optimally doped systems, and were considered as a starting point for the calculations: in solving the Eliashberg equations, parameters were adjusted to give the experimental $T_c$ values. Concerning the energy gaps, in the calculation we fixed the input parameters so that the calculated gaps would reproduce as initial low-temperature values those reported in literature for ARPES measurements, when available \cite{Ding2008,Inosov2011,Mansart2012,Yoshida2014,Ye2012}, but then we let them change to better reproduce our experimental data. At the end of the procedure, a good agreement with ARPES measurements still holds. In Fig.\ref{Fig_Delta} the imaginary axis solutions for $\Delta_i(T)$ are shown as a function of temperature, while in Table \ref{tab:parametri} we reported the low-temperature values $\Delta^R_i(0)$ on the real axis (that coincide with values obtained as analytical continuations from the imaginary axis by Pad\'e approximants).  

\textit{(b) Penetration depth} $-$ The temperature dependencies of the penetration depth for the three compounds, shown in Fig.\ref{Fig_lambda} in the form of $\Delta\lambda_L=\lambda_L(T)-\lambda_L(0)$, are consistent with literature \cite{Gordon2009}. As for the absolute values, accessible to our experiment ($\lambda_L(0)$ reported in Table \ref{tab:parametri}), literature data are rare, but still the agreement holds considering that values around 200 nm are usually reported for $\lambda_L(0)$ of Ba122 compounds \cite{Gordon2010}. The agreement between experimental curves and calculated ones is excellent, if the temperature dependence is considered (see the comparison in Fig.\ref{Fig_lambda}), but a distinction has to be made from the K-substituted system, for which the matching is valid also for the absolute values, to the P- and Co-substituted compounds, that show a relevant deviation of the calculated $\lambda_L(0)$ with respect to the experimental one, as mentioned above. This has generally been

\begin{figure}[t]
\begin{center}
\includegraphics[keepaspectratio, width=\columnwidth]{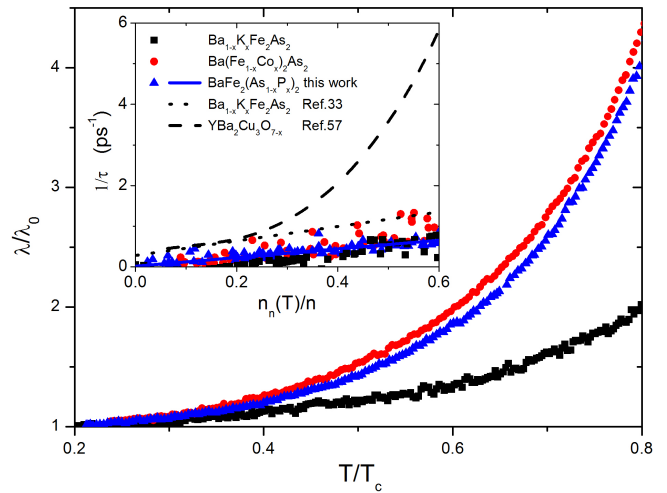}
\caption{Experimental $\lambda_L(T)/\lambda_L(0)$ vs reduced temperature, $T/T_c$ for the three systems. The inset shows the quasiparticle scattering rate $1/\tau$ as a function of the normalized quasiparticle density $n_n(T)/n$ for the IBSs under investigation, with a comparison to the results in Ba$_{1-x}$K$_{x}$Fe$_{2}$As$_{2}$ at 28 GHz (dotted line)\cite{Hashimoto2009PRL2} and the results in YBa$_2$Cu$_3$O$_{6.95}$ at 34.8 GHz (dashed line)\cite{Bonn}.}\label{Fig_nodi}
\end{center}
\end{figure}

\begin{widetext}
\begin{center}
\begin{table}[h!]
\caption{The Table summarizes, for the three different dopants under study, the values of the experimental outputs and of the main model parameters used to reproduce the experimental data. $T_c$ is the experimental critical temperature, $\lambda_L^{exp}(0)$ is the low-temperature penetration depth, and $\sigma(T_c)$ is the microwave conductivity at $T_c$. Concerning the theoretical and model parameters, $\lambda_{ij}$ are the components of the electron-boson coupling-constant matrix, $\Delta_i^R(0)$ are the low-temperature values of the gaps on the real axis, $\omega_p/2\pi$ is the plasma frequency, $w^\lambda_{i}$ and $w^\sigma_{i}$ are the weights of the $i$-th band in equations \ref{eq.lambda} and \ref{eq.cond}, respectively.  \label{tab:parametri}}
\renewcommand{\arraystretch}{1.5}
\begin{tabular}{p{1cm} p{0.8cm} p{1.1cm} p{1.5cm} p{0.8cm} p{0.8cm} p{0.8cm} p{0.8cm} p{1cm} p{1cm} p{1cm} p{1cm} p{2.2cm} p{2.2cm}}
\hline
\hline
\makebox[1cm][c]{dopant} & \makebox[0.8cm][c]{$T_c$} & \makebox[1.1cm][c]{$\lambda_L^{exp}(0)$}  & \makebox[1.5cm][c]{$\sigma(T_c)$} & \makebox[0.8cm][c]{$\lambda_{12}$} & \makebox[0.8cm][c]{$\lambda_{23}$} & \makebox[0.8cm][c]{$\lambda_{13}$}  & \makebox[0.8cm][c]{$\lambda_{ii}$} & \makebox[1cm][c]{$\Delta^R_1(0)$} & \makebox[1cm][c]{$\Delta^R_2(0)$}  & \makebox[1cm][c]{$\Delta^R_3(0)$}  & \makebox[1cm][c]{$\hbar\omega_p$} & \makebox[2.2cm][c]{$w^\lambda_{1,2,3}$} & \makebox[2.2cm][c]{$w^\sigma_{1,2,3}$}\\
\makebox[1cm][c]{ } & \makebox[0.8cm][c]{(K)} & \makebox[1.1cm][c]{(nm)}  & \makebox[1.5cm][c]{($\Omega^{-1}m^{-1}$)} & \makebox[0.8cm][c]{ } & \makebox[0.8cm][c]{ } & \makebox[0.8cm][c]{ }  & \makebox[0.8cm][c]{ } & \makebox[1cm][c]{(meV)} & \makebox[1cm][c]{(meV)}  & \makebox[1cm][c]{(meV)}  & \makebox[1cm][c]{(meV)} & \makebox[2.2cm][c]{} & \makebox[2.2cm][c]{}\\
\hline
\hline
\makebox[1cm][c]{K}& \makebox[0.8cm][c]{38.7} & \makebox[1.1cm][c]{197}  & \makebox[1.5cm][c]{1.95$\cdot 10^{6}$} & \makebox[0.8cm][c]{0.00} & \makebox[0.8cm][c]{0.75} & \makebox[0.8cm][c]{3.37}  & \makebox[0.8cm][c]{0.00} & \makebox[1cm][c]{12.0} & \makebox[1cm][c]{5.4}  & \makebox[1cm][c]{-12.0}  & \makebox[1cm][c]{1.00} & \makebox[2.2cm][c]{0.1, 0.1, 0.8} & \makebox[2.2cm][c]{0.6, 0.2, 0.2}\\
\hline 
\makebox[1cm][c]{Co}& \makebox[0.8cm][c]{24.2} & \makebox[1.1cm][c]{165}  & \makebox[1.5cm][c]{0.47$\cdot 10^{6}$} & \makebox[0.8cm][c]{0.20} & \makebox[0.8cm][c]{0.00} & \makebox[0.8cm][c]{2.72}  & \makebox[0.8cm][c]{0.30} & \makebox[1cm][c]{7.2} & \makebox[1cm][c]{-3.9}  & \makebox[1cm][c]{-7.8}  & \makebox[1cm][c]{0.20} & \makebox[2.2cm][c]{0.85, 0.05, 0.1} & \makebox[2.2cm][c]{0.39, 0.26, 0.35}\\ 
\hline 
\makebox[1cm][c]{P}& \makebox[0.8cm][c]{29.0} & \makebox[1.1cm][c]{160}  & \makebox[1.5cm][c]{1.13$\cdot 10^{6}$} & \makebox[0.8cm][c]{0.00} & \makebox[0.8cm][c]{7.69} & \makebox[0.8cm][c]{0.70}  & \makebox[0.8cm][c]{0.00} & \makebox[1cm][c]{3.8} & \makebox[1cm][c]{10.8}  & \makebox[1cm][c]{-8.3}  & \makebox[1cm][c]{0.55} & \makebox[2.2cm][c]{0.5 ,0.4, 0.1} & \makebox[2.2cm][c]{0.12, 0.53, 0.35}\\
\hline
\hline
\end{tabular} 
\end{table}
\end{center}
\end{widetext}

\noindent ascribed to Fermi-liquid effects, not taken into account by the theory \cite{Vorontsov2009}, but we argue that these explanations alone are not satisfactory, since strong coupling effects are accounted by the renormalization functions $Z_{i}(i\omega_{n})$ in Eliashberg models. The origin of this discrepancy still is not well understood and it is worthy to be further investigated. Possibly, it is due to the fact that the adopted relation between $\lambda_L(0)$ and the plasma frequency loses its validity (this is the reason why, as a reference, we preferred to list the calculated $\omega_p$ in Table \ref{tab:parametri}). A possible reason for this could lie in the fact that vertex corrections are not accounted for in our model. In order to further compare the compounds, we plot in Fig.\ref{Fig_nodi} experimental $\lambda_L(T)/\lambda_L(0)$ vs reduced temperature, $T/T_c$. We notice that the curve of the K-doped system is qualitatively different from the other much steeper ones.  

We point out that the main difference between P- and Co-doping with respect to K-doping lies in the fact that in the former the chemical substitution is performed on the FeAs planes, that are the main responsible for the superconducting properties. Moreover, it has been suggested from the spin fluctuations theory that the three-dimensionality of the Fermi surfaces (and thus the superconducting gap structure) is quite sensitive to the pnictogen position (height from the FeAs plane) \cite{Kasahara2010}. The substitution of P for As reduces this pnictogen height, which is not the case for the hole doping by K substitution for Ba \cite{Hashimoto2010PRB}. Thus, we suggest that the observed peculiar behaviours, i.e. the $\lambda_L(T)/\lambda_L(0)$ curves and the anomalous $\lambda_L(0)-\omega_p$ relation, could be ascribed to the fact that chemical disorder is introduced directly in the FeAs superconducting planes and/or the height of the pnictogen from the planes is varied. It should be considered that this disorder could also originate scattering, but $-$ as argued in Ref.\cite{Ghigo2017prb} $-$ in our calculations we need to consider the pristine optimally doped compound as ideal, setting to zero all the scattering rates, otherwise the number of possible free parameters in the system becomes too large to be treated.

\textit{(c) Quasiparticle conductivity} $-$ It is important to discuss the temperature dependence of the quasiparticle conductivity shown in Fig.\ref{Fig_sigma}, because its features could give valuable information about the symmetry of the superconducting order parameter. In particular, the observation of the so-called coherence peak in the ac conductivity is usually taken as evidence for a uniform gap function, although more precisely it is an evidence that the portions of the Fermi surface coupled by the experimental probe have gaps of the same sign and similar magnitude \cite{Aguilar2010}. It is therefore important to distinguish between techniques that probe the system with small or large momentum transfer. In a dirty isotropic $s$-wave BCS superconductor the coherence peak appears in the quasiparticle conductivity slightly below $T_c$ \cite{klein1994}. On the contrary, in cuprate superconductors no coherence peak just below $T_c$ is observed, but rather a peak which can have a larger amplitude is seen much below the critical temperature \cite{Deepwell2013}. It is usually ascribed to the concomitant decrease of the quasiparticle density and the increase of the quasiparticle scattering time, upon cooling. Which is the case for IBS, and for Ba122 systems in particular, is still under debate. The emergence of a coherence peak in these compounds is also a matter of the used probe: it was not observed by NMR \cite{Yu2011,Yashima2009}, because it is a local probe and can couple parts of the Fermi surface that differ by large momentum transfer. Thus, the absence of the coherence peak in the NMR relaxation rate has been interpreted as supporting the picture of the sign changing extended $s$-wave symmetry of the gap function: it can be suppressed by a partial cancellation of total susceptibility, owing to the sign change between the hole and electron bands \cite{Hashimoto2009PRL2}. On the other hand, the coherence peak have been observed by terahertz conductivity measurements on Ba(Fe$_{1-x}$Co$_x$)$_{2}$As$_{2}$ \cite{Aguilar2010,Fischer2010}: in this case, since at terahertz frequencies the photon small momentum allows one to probe only zero-momentum excitations around the Fermi surface, only a single sign of the order parameter is detected because different sheets are separated by large momentum transfer. Thus a coherence peak qualitatively resembling that from a single-uniform-gap superconductor can be measured. The same consideration holds for conductivity measurements at microwaves, due to the long wavelength. Indeed, a clue of a coherence peak was detected by microwave conductivity measurements in Ba$_{1-x}$K$_{x}$Fe$_{2}$As$_{2}$, even in the presence of another larger and higher peak at lower temperatures \cite{Hashimoto2009PRL2}. This last peak, due to the same mechanism described above for the cuprates, was observed also in other IBS systems, and sometimes it completely masks possible coherence peaks \cite{Takahashi2011}. As a matter of facts, a trace of the coherence peak can be better highlighted after subtraction of a residual surface resistance term from data \cite{Ghigo2016}. Here, we show on bare data without any specific treatment that the quasiparticle conductivity can be well described by our model $-$ implying s$_\pm$ symmetry $-$ with the same parameters already adjusted to fit the penetration depth curves. Figure \ref{Fig_sigma} shows that the model succeeds to capture the main experimental trend, specifically the broad peak below $T_c/2$, thus confirming its validity. Nevertheless, a smaller contribution to the experimental curves of a coherence peak cannot be excluded.

\textit{(d) Nodes in the energy gap} $-$ Another item that could be a cause of different behaviours of these compounds is the possible presence of nodes in the energy gap. In fact, although the Fermi-surface topology is similar in these systems, slight differences in the size and corrugation of hole surfaces may give rise to dramatic changes in the nodal topology \cite{Hashimoto2010PRB}. Experimental evidences for line nodes in BaFe$_{2}$(As$_{1-x}$P$_x$)$_{2}$ at optimal doping was claimed by Nakai \textit{et al.} (NMR) \cite{nakai2010}  and by Yamashita \textit{et al.} (angle-resolved thermal conductivity) \cite{Yamashita}. Other observations seem to suggest the tendency of going from a fully gapped system for x=0.32 to a system with pronounced gap anisotropy and possible nodes for overdoped x=0.55 (specific heat) \cite{Diao}. ARPES data by Yoshida \textit{et al.} are inconsistent with horizontal nodes but are consistent with modified s$_\pm$ gap with nodal loops \cite{Yoshida2014}. A circular line node on the largest hole Fermi surface was found by Zhang \textit{et al.} by ARPES, ruling out a $d$-wave pairing origin for the nodal gap, and establishing the existence of nodes in IBS under the $s$-wave pairing symmetry \cite{Zhang2012}. This ring node would not be forced by symmetry, but rather it should be an "accidental" one.\\
An alternative way to check if the order parameter is nodeless or not, more connected to the present work, is the analysis of the temperature dependence of the penetration depth. It was reported that, for a pure superconductor in a $d$-wave state at temperatures well below $T_c$, $\lambda$ should show a $T-$linear dependence \cite{Hirschfeld1993}. Reversing this argument, Hashimoto \textit{et al.} stated that the linear $\lambda(T)$ they observed at low temperatures in BaFe$_{2}$(As$_{1-x}$P$_x$)$_{2}$ testifies that this system presents $d$-wave-like line nodes \cite{Hashimoto2010PRB,Hashimoto_Science}. However, this behavior is only shown in a 3-4 K range below $T/T_c=0.15$, not accessible by our experiment.

Alternatively, one can plot the quasiparticle scattering rate $1/\tau$ as a function of the normalized quasiparticle density $n_n/n$ (where the quasiparticle density is $n_n(T)=n-n_s(T)$). In an $s$-wave superconductor without nodes, a linear relation between $1/\tau(T)$ and $n_n(T)$ is expected \cite{Hashimoto2009PRL2}, while in a $d$-wave superconductor it is superlinear ($1/\tau \sim n_n^3$). In the inset of Fig.\ref{Fig_nodi} we plotted data of the IBSs under investigation, with comparisons to the results reported in Ba$_{1-x}$K$_{x}$Fe$_{2}$As$_{2}$ at 28 GHz by Hashimoto \textit{et al.} \cite{Hashimoto2009PRL2} and results in the $d$-wave YBa$_2$Cu$_3$O$_{6.95}$ at 34.8 GHz \cite{Bonn}. All the Ba122 systems show a linear trend, similar to that of Ref.\cite{Hashimoto2009PRL2}, very different from the superlinear behavior of YBa$_2$Cu$_3$O$_{6.95}$. Thus, from our data the presence of $d$-wave-like node lines in BaFe$_{2}$(As$_{1-x}$P$_x$)$_{2}$ seems to be excluded, however our data does not exclude (nor suggest) other nodal structures, e.g. nodes loops, vertical line nodes and other accidental nodes that are consistent with $s$-wave symmetry that have been observed\cite{Yoshida2014} and explained theoretically\cite{Saito2013}. Whether this depends on the experimental technique, or on the nature of the nodes itself deserves further studies and is currently under investigation.

\textit{Conclusions} $-$ In summary, we have shown that an approach based on a three band, s$_\pm$-wave Eliashberg model is able to explain in a self-consistent way a set of experimental data ranging from the critical temperature, to the penetration depth and to the microwave conductivity of different Ba122 optimally doped single crystals, with different substitutions.\\ 
This remarkable result allowed us to discuss details of the observed behaviour of IBS single crystals and their connection to structural properties of the samples and to the symmetry of the superconducting order parameter. We suggest that a relevant role in the variation of such properties, additional to the effects of charge doping alone, could be played by chemical substitution in the FeAs superconducting planes. The behaviour of the quasiparticle conductivity can be explained by the model without subtraction of a residual surface resistance contribution. In particular, a wide peak observed below $T_c/2$ can be understood from the temperature dependence of the scattering time, while it probably masks the existence of coherence effects that could in principle emerge in a measurement that uses a small momentum probe. At last, our data seem to rule out the existence of $d$-wave-like line nodes, but could be consistent with other nodal structures (such as accidental nodal loops or vertical line nodes).

\begin{acknowledgments}
This work is partly supported by KAKENHI (Grant No. 17H01141) from JSPS. G.A.U. acknowledges support from the MEPhI Academic Excellence Project (Contract No. 02.a03.21.0005)
\end{acknowledgments}

\end{document}